\documentstyle[emulateapj]{article}

\slugcomment{Submitted to the Astrophysical Journal}

\received{2002 September 3}
\begin{document}

\title{X-ray Spectral Properties of Low-Mass X-ray
Binaries in Nearby Galaxies}

\author{Jimmy A. Irwin\altaffilmark{1,2}, Alex E. Athey\altaffilmark{1},
Joel N. Bregman\altaffilmark{1}}

\altaffiltext{1}{Department of Astronomy, University of Michigan,
Ann Arbor, MI 48109-1090
E-mail: jirwin@astro.lsa.umich.edu, aathey@umich.edu, jbregman@umich.edu}

\altaffiltext{2}{{\it Chandra} Fellow.}

\begin{abstract}
We have investigated the X-ray spectral properties of a collection of
low-mass X-ray binaries (LMXBs) within a sample of 15 nearby
early-type galaxies using proprietary and archival data from the {\it Chandra
X-ray Observatory}. We find that the spectrum of the sum of the sources in a
given galaxy is remarkably similar from galaxy to galaxy when only sources
with X-ray luminosities less than $10^{39}$ ergs s$^{-1}$ (0.3--10 keV) are
considered. Fitting these lower luminosity sources in all galaxies
simultaneously with a power law model led to a best-fit power law exponent of
$\Gamma = 1.56 \pm 0.02$ (90\% confidence), and using a thermal bremsstrahlung
model yielded $kT_{brem} = 7.3 \pm 0.3$ keV. This is the tightest
constraint to date on the spectral properties of LMXBs in external galaxies.
The spectral properties of the LMXBs do not vary with
galactic radius out to three effective radii. There is also no apparent
difference in the spectral properties of LMXBs that reside within globular
clusters and those that do not. We demonstrate how the
uniformity of the spectral properties of LMXBs can lead to more accurate
determinations of the temperature and metallicity of the hot gas in galaxies
that have comparable amounts of X-ray emission from hot gas and LMXBs.

Although few in number in any given galaxy, sources with luminosities of
$1-2 \times 10^{39}$ ergs s$^{-1}$ are present in 10 of the galaxies.
The spectra of these luminous sources are softer than the
spectra of the rest of the sources, and are consistent with the spectra
of Galactic black hole X-ray binary candidates when they are in their
very high state. The spatial distribution of these sources is much flatter
than the optical light distribution, suggesting that a significant portion of
them must reside within globular clusters. The simplest explanation of these
sources is that they are $\sim 10-15$ M$_{\odot}$ black holes accreting near
their Eddington limit. The spectra of these sources are very different than
those of ultraluminous X-ray sources (ULXs) that have been found within
spiral galaxies, suggesting that the two populations of X-ray luminous
objects have different formation mechanisms.
The number of sources with apparent luminosities above
$2 \times10^{39}$ ergs s$^{-1}$ when determined using the distance of the
galaxy is equal to the number of expected background AGN and thus appear
to not be associated with the galaxy, indicating that very luminous sources
are absent or very rare in early-type galaxies. The lack of
ULXs within elliptical galaxies strengthens the argument that ULXs
are associated with recent star formation.
\end{abstract}
\keywords{binaries: close --- X-rays: galaxies --- X-rays: stars}


\section{Introduction} \label{sec:intro}

The {\it Chandra X-ray Observatory} has made it possible to resolve
dozens if not hundreds of individual X-ray point sources in nearby galaxies
owing to its sub-arcsecond spatial resolution (Sarazin, Irwin, \& Bregman
2000; Angelini, Loewenstein, \& Mushotzky 2001; Kraft et al.\ 2001;
Bauer et al.\ 2001; Soria \& Wu 2002). While spiral galaxies contain
a variety of types of X-ray point sources (high- and low-mass X-ray binaries,
supernovae remnants), elliptical and S0 galaxies, as well as the bulges
of spiral galaxies contain almost exclusively low-mass X-ray binaries (LMXBs).
LMXBs are believed to be composed of a compact accreting primary (a neutron
star or a black hole) and a low-mass main sequence or red giant secondary that
is losing material to the primary as a result of Roche lobe overflow.
Although LMXBs have been studied extensively in our Galaxy, resolving them
and determining their spatial distribution and spectral characteristics
in external galaxies has only recently become feasible.

\begin{table*}[t]
\scriptsize
\caption{Sample of Galaxies}
\label{tab:sample}
\begin{center}
\begin{tabular}{lcccccccccc}
\multicolumn{10}{c}{} \cr
\tableline \tableline
Galaxy & Galaxy &Type & Obs. & Distance & Semimajor & Semiminor &
$N_H$ & Exposure & Luminosity \cr
Number &&&ID & (Mpc) & Axis ($^{\prime\prime}$) & Axis ($^{\prime\prime}$) &
($10^{20}$ cm$^{-2}$) & (seconds) &
Limit (ergs s$^{-1}$)\cr
\tableline
\phn\phn1 & NGC~1291 & Sa & 795 & 8.9 & \ldots  & \ldots &2.12 & 37,406 &
$2.0 \times 10^{37}$  \\
\phn\phn2 & NGC~1316 & S0 & 2022 & 21.5 & 132.2\tablenotemark{a} &
90.5\tablenotemark{a} & 1.88 & 24,478 & $1.7 \times 10^{38}$  \\
\phn\phn3 & NGC~1399 & E1 & 319 & 20.0 & 44.6\tablenotemark{b} &
40.5\tablenotemark{b} & 1.34 & 54,540 & $5.7 \times 10^{37}$   \\
\phn\phn4 & NGC~1407 & E0 & 791 & 28.8 & 73.9\tablenotemark{b} &
68.7\tablenotemark{b} & 5.42 & 33,763 & $2.8 \times 10^{38}$   \\
\phn\phn5 & NGC~1549 & E0 & 2077 & 19.7 & 51.0\tablenotemark{b} &
44.7\tablenotemark{b} & 1.46 & 21,892 & $1.8 \times 10^{38}$    \\
\phn\phn6 & NGC~1553 & S0 & 783 & 18.5 & 78\tablenotemark{c} &
51\tablenotemark{c} & 1.50 & 16,361 & $1.6 \times 10^{38}$    \\
\phn\phn7 & NGC~3115 & S0 & 2040 & 9.7 & 93\tablenotemark{d} &
35\tablenotemark{d} & 4.32 & 36,979 & $2.6 \times 10^{37}$    \\
\phn\phn8 & NGC~3585 & E/S0 & 2078 & 20.0 & 56\tablenotemark{e} &
28\tablenotemark{e} & 5.57 & 33,706 & $1.7 \times 10^{38}$   \\
\phn\phn9 & NGC~4374 & E1 & 803 & 18.4 &58.2\tablenotemark{b} &
53.3\tablenotemark{b} &  2.60 & 28,049 & $9.8 \times 10^{37}$   \\
\phn\phn10 & NGC~4472 & E2 & 321 & 16.3 & 114.0\tablenotemark{f} &
95.6\tablenotemark{f} & 1.66 & 26,326 & $9.5 \times 10^{37}$    \\
\phn\phn11 & NGC~4494 & E1 & 2079 & 17.1 & 49.3\tablenotemark{b} &
42.2\tablenotemark{b} & 1.52 & 18,254 & $1.3 \times 10^{38}$    \\
\phn\phn12 & NGC~4636 & E/S0 & 323 & 14.7 & 117.0\tablenotemark{f} &
85.6\tablenotemark{f} & 1.81 & 42,686 & $5.2 \times 10^{37}$   \\
\phn\phn13 & NGC~4649 & E2 & 785 & 16.8 & 82.0\tablenotemark{f} &
66.2\tablenotemark{f} & 2.20 & 18,401 & $1.4 \times 10^{38}$    \\
\phn\phn14 & NGC~4697 & E6 & 784 & 11.8 & 97.3\tablenotemark{b} &
60.2\tablenotemark{b} & 2.12 & 39,063 & $3.1 \times 10^{37}$    \\
\phn\phn15 & M31 & Sb & 309 & 0.76 & \ldots & \ldots & 6.66 & 5,089 &
$1.2 \times 10^{36}$    \\

\tableline
\end{tabular}
\end{center}
\tablenotetext{a}{Caon, Capaccioli, \& D'Onofrio (1994)}
\tablenotetext{b}{Goudfrooij et al.\ (1994)}
\tablenotetext{c}{Kormendy (1984) and Jorgensen, Franx, \& Kjaergaard (1995)}
\tablenotetext{d}{Capaccioli, Held, \& Nieto (1987)}
\tablenotetext{e}{Ryden, Forbes, \& Terlevich (2001)}
\tablenotetext{f}{Peletier et al.\ (1990)}

\end{table*}

Initial work on the LMXB population of early-type galaxies has led to several
interesting results. Sarazin, Irwin, \& Bregman (2000, 2001) detected 90 X-ray
sources in a {\it Chandra} observation of the elliptical galaxy NGC~4697,
and found a break in the luminosity function of the sources at a luminosity of
$\sim3 \times 10^{38}$ ergs s$^{-1}$, intriguingly close to the
Eddington luminosity of an accreting 1.4 M$_{\odot}$ neutron star. One
interpretation of this break is that it represents a division between
black hole X-ray binaries and neutron star X-ray binaries, with only black hole
binaries more luminous than the break, and a mixture of neutron star
and black hole binaries less luminous than the break. Such breaks have
also been found in other early-type galaxies (Blanton, Sarazin, \& Irwin 2001;
Finoguenov \& Jones 2002;
Randall, Sarazin, \& Irwin 2002; Kundu, Maccarone, \& Zepf 2002). Recent
studies have also found that a significant fraction (40\%--70\%) of the
X-rays binaries reside in globular clusters of the host galaxy
(Angelini et al.\ 2001; Randall et al.\ 2002;
Kundu et al.\ 2002), in marked contrast to the $\sim10\%$ of Galactic and M31
LMXBs that reside within globular clusters. In addition, when the spectra of
all the resolved LMXBs of a given galaxy were added together and fit with a
power law spectral model, an index ranging from 1.5--2.0 has been obtained
(Sarazin et al.\ 2001; Randall et al.\ 2001; Kim \& Fabbiano 2002;
Irwin, Sarazin, \& Bregman 2002).

Previous X-ray satellites lacked the needed spatial resolution and bandpass
coverage to separate cleanly the LMXB component from the hot gas
component in early-type galaxies. While recent progress has been made from
the study of the luminosity functions and host
environment of LMXBs in early-type galaxies, a detailed analysis of the
spectral properties of LMXBs with a large sample of galaxies has not yet
been attempted with {\it Chandra}.
With {\it Chandra} we are now in a position to determine the spectral
properties of individual bright sources in galaxies or add up the spectra
of the fainter sources to determine the bulk spectral properties of the LMXBs.

A more thorough understanding of the X-ray spectral properties of LMXBs in
early-type galaxies is critical for the study of the hot, X-ray--emitting
gas within these systems. Although the LMXB contribution to the X-ray emission
from gas-rich galaxies is negligible, this is not the case for galaxies with
moderate to low amounts of X-ray--emitting gas. In these galaxies, the LMXB
contribution can be the dominant X-ray emission mechanism. Quantifying the
spectral properties of LMXBs in nearby galaxies where the LMXBs are
resolvable will allow for a more accurate separation of the gaseous and
stellar X-ray components in galaxies too distant for the individual LMXBs
to be detected.

In this paper we use both proprietary and archival {\it Chandra}
data for 15 early-type systems (consisting of eight elliptical galaxies, two
transitional E/S0 galaxies, three S0 galaxies, and two spiral bulges) to
determine the spectral characteristics of LMXBs over a range of X-ray
luminosity classes, as well as a function of galactic
radius. Unless otherwise stated, all uncertainties are 90\% confidence
levels, and all X-ray luminosities are in the 0.3--10 keV energy band.

\section{Observations and Data Reduction} \label{sec:observations}

We have constructed a sample of 13 early-type galaxies and two spiral bulges
that were observed with the ACIS S3 chip onboard {\it Chandra}. The sample is
given in Table~\ref{tab:sample}. The assumed distances were taken from
Tonry et al.\ (2001) with the exceptions of NGC~1291 (de Vaucouleurs 1975)
and M31 (van den Bergh 2000), and the assumed Galactic hydrogen column
densities are from Dickey \& Lockman (1990).

The galaxies in the sample were processed in a uniform manner following the 
{\it Chandra} data reduction threads employing CIAO 2.2.1 coupled with
CALDB 2.12. All of the 
data were calibrated with the most recent gain maps at the time of reduction
(acisD2000-08-12gainN0003.fits for --120$^\circ$ C focal plane temperature data
and acisD2000-08-12gainN0003.fits for --110$^\circ$ C data).
{\it ASCA} grades of 0, 2, 3, 4 and 6 were selected for all subsequent 
processing and analysis. The observation specific bad pixel files were applied
from the standard calibration library included with CIAO 2.2.1. Pile-up
was not an issue even for the brightest sources and no correction has
been applied.

To check for background flares, a temporal light curve was constructed from the
outer regions of the ACIS S3 chip. We examined each light curve by eye and
eliminated any data approximately three to four sigma away from a determined
mean value during a quiescent period. Typically this eliminated less than
15\% of the data. Since our goal is to detect point sources rather than diffuse
emission, strict screening of high background rates is not necessary.
The peak of the energy distribution of the source region
was determined and used to generate an exposure map appropriate for that
energy for each observation. Sources were detected using the ``Mexican-Hat"
wavelet detection routine {\sc wavdetect} in CIAO in an 0.3--6.0 keV
band image. The threshold was set to give approximately one false detection per
image and the scales run were from 2--32 pixels. The source list of each galaxy
was culled to remove any source not detected at the 3$\sigma$ level.

In order to reduce contamination from unrelated foreground/background
X-ray sources, we restrict our study to sources within three effective
(half-optical light) radii
of the center of each galaxy. Effective semimajor and semiminor axes
as well as position angles were collected from the literature (see
Table~\ref{tab:sample}).
In galaxies
with large amounts of hot, diffuse gas (i.e., NGC~1399, NGC~4472), some
sources near the center of the galaxy were omitted since their count rates
were highly uncertain owing to the presence of the high X-ray surface
brightness gas.

For each source a local background was determined from a circular annular
region with an inner radius that was set to 1.5 times the semimajor axis
of the source extraction region and an outer
radius chosen such that the area of the background annulus was five
times the area of the sources extraction region. Care was taken to exclude
neighboring sources from the background annuli of each source in crowded
regions. We excluded point sources located at the optical center of each
galaxy since it is likely that many of these sources are actually
low-luminosity active galactic nuclei (AGN) rather than LMXBs.

\section{Spectral Analysis} \label{sec:spectral}

For each galaxy, the spectra of all the sources in the galaxy within three
effective radii were extracted and summed into one spectrum using the
CIAO routine {\sc acisspec}. This routine calculates a weighted
redistribution matrix file (RMF) and ancillary response file (ARF)
appropriate for extended sources, or for our case, a collection of
point sources spread out over the ACIS detector. The ARF files have
been corrected for the continuous degradation of the ACIS quantum
efficiency\footnote{See
http://cxc.harvard.edu/cal/Links/Acis/acis/Cal\_prods/qeDeg/index.html for a
complete description.} using the CIAO tool {\sc corrarf}, which applies
the {\sc ACISABS} absorption profile (Chartas \& Getman 2002\footnote{See
http://www.astro.psu.edu/users/chartas/xcontdir/xcont.html.})
to the original ARF
file. The composite spectrum was then regrouped such that each energy channel
contained at least 25 X-ray counts. Channels with energies less than 0.5 keV
and greater than 6.0 keV were then excluded. The composite spectrum for
each galaxy was then fit within XSPEC v11.2.0 with a simple power law
model absorbed by the Galactic hydrogen column density
(Table~\ref{tab:sample}).

The results of these fits are shown in Figure~\ref{fig:all}
for each of the 15 galaxies in our sample.
The best-fit power law exponent varies from  1.45 to 1.9 from galaxy
to galaxy, in agreement with previous {\it Chandra} studies of LMXBs in
early-type systems. In general, freeing the absorption column density did
not improve the quality of the fits, so the column densities have remained
fixed at the Galactic value throughout the remainder of this paper. For the
individual sources in each particular galaxy, the best-fit power law exponent
for that galaxy was used to convert the source count rates into
0.3--10 keV luminosities.
The correction for the degradation of the ACIS quantum efficiency typically
increased the power law exponent by 10\%--20\% over the uncorrected value,
with the correction greatest for the most recently observed targets, as is
expected for linearly increasing absorption over time from
molecular contamination of the ACIS optical blocking filters.
\bigskip

\centerline{\null}
\vskip3.30truein
\includegraphics{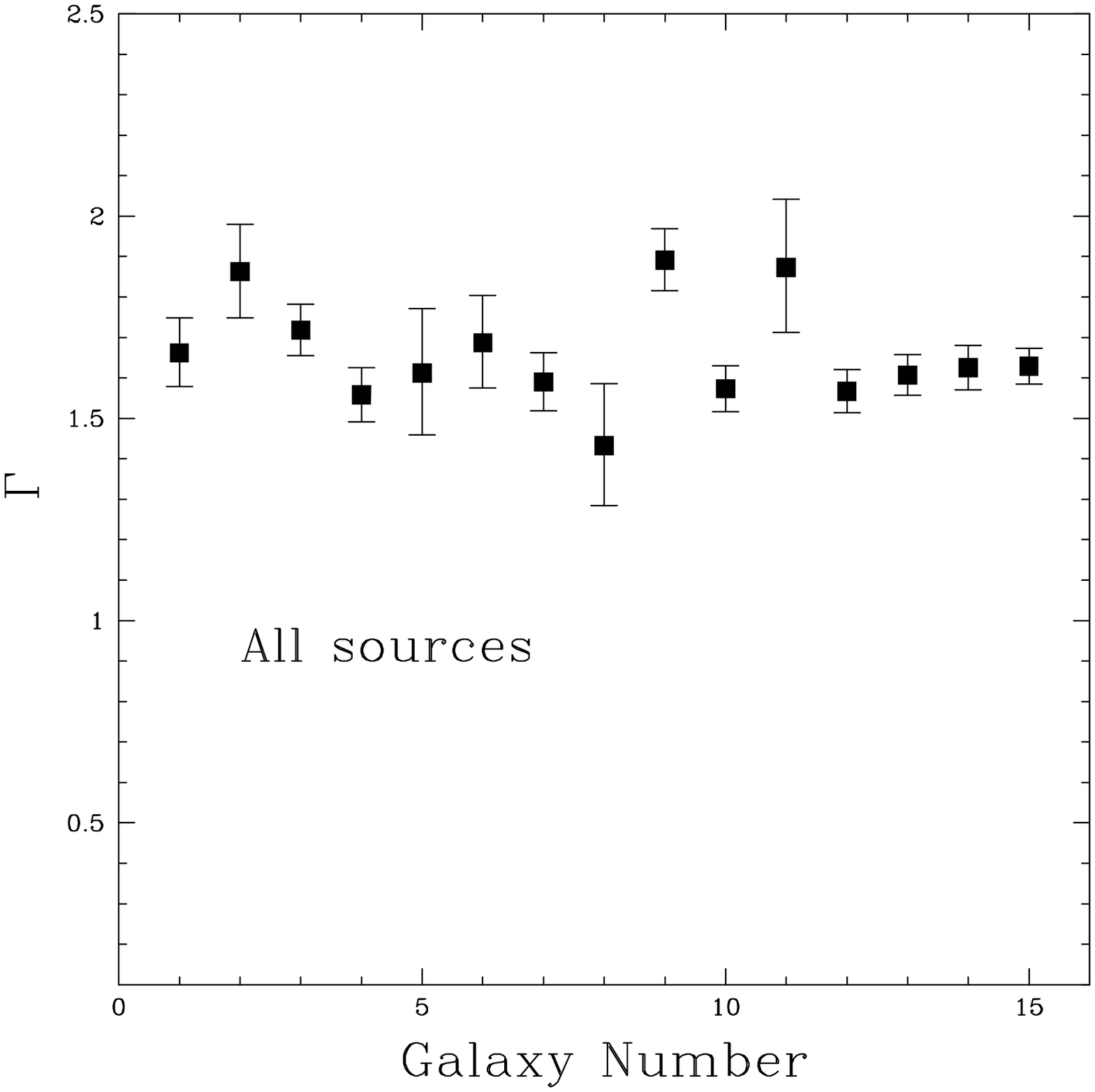}
\figcaption{\small
Best-fit power law exponent ($\Gamma$) when the spectra of the sources were
summed for each galaxy in our sample. The galaxy number corresponds to
the numbering system in Table~\protect\ref{tab:sample}.
\label{fig:all}}
\bigskip
Next, we generated color-color plots for the individual sources in each galaxy.
We define three X-ray bands, S (0.3--1.0 keV),
M (1.0--2.0 keV), and H (2.0--6.0 keV), and create two X-ray colors,
H21 = (M--S)/(M+S) and H31 = (H--S)/(H+S). The color-color plots of the
15 galaxies in our sample made it apparent that the most luminous sources
($L_X > 10^{39}$ ergs s$^{-1}$) tend to have significantly softer colors than
the rest of the sources. This is illustrated in Figure~\ref{fig:colors} where
the H21 vs.\ H31 values have been plotted for the sources within all the
galaxies except NGC~1407, NGC~3115, NGC~3585, and M31, which have higher
Galactic column densities that will artificially harden the colors.
It is clear that the most luminous sources are indeed softer
on average than the less luminous sources. We have performed a two-dimensional
Kolmogorov-Smirnov test on the faint and bright sources and found that the
probability that the H21 and H31 colors of the two distributions were chosen
from the same parent distribution was only 0.004.

Also note the presence of
a small number of sources at (--1, --1). These sources have no detectable
emission above 1 keV and are most likely supersoft sources, such as those
seen in the Milky Way and the disk of M31. The traditional explanation of
these supersoft sources is that they are accreting white dwarfs burning
hydrogen on their surface (van den Heuvel et al.\ 1992). However, the
bolometric luminosity of such a supersoft in M81 far exceeds the
Eddington limit for a 1.4 M$_{\odot}$ white dwarf, prompting
Swartz et al.\ (2002) to postulate that these sources are intermediate
mass black holes ($10^2 - 10^3$  M$_\odot$) accreting at a few percent
of their Eddington limit, rather than a white dwarf. The brighter supersoft
sources in our sample have similar bolometric luminosities. However, the
small number of these sources found in our sample combined with their
low count rates in the 0.5--6 keV energy range limits what we can learn
from them, so we have excluded them from our analysis in order to focus
on LMXBs. A thorough study of supersoft sources in less distant galaxies
has recently been conducted by Di Stefano \& Kong (2002).

\centerline{\null}
\vskip3.50truein
\includegraphics{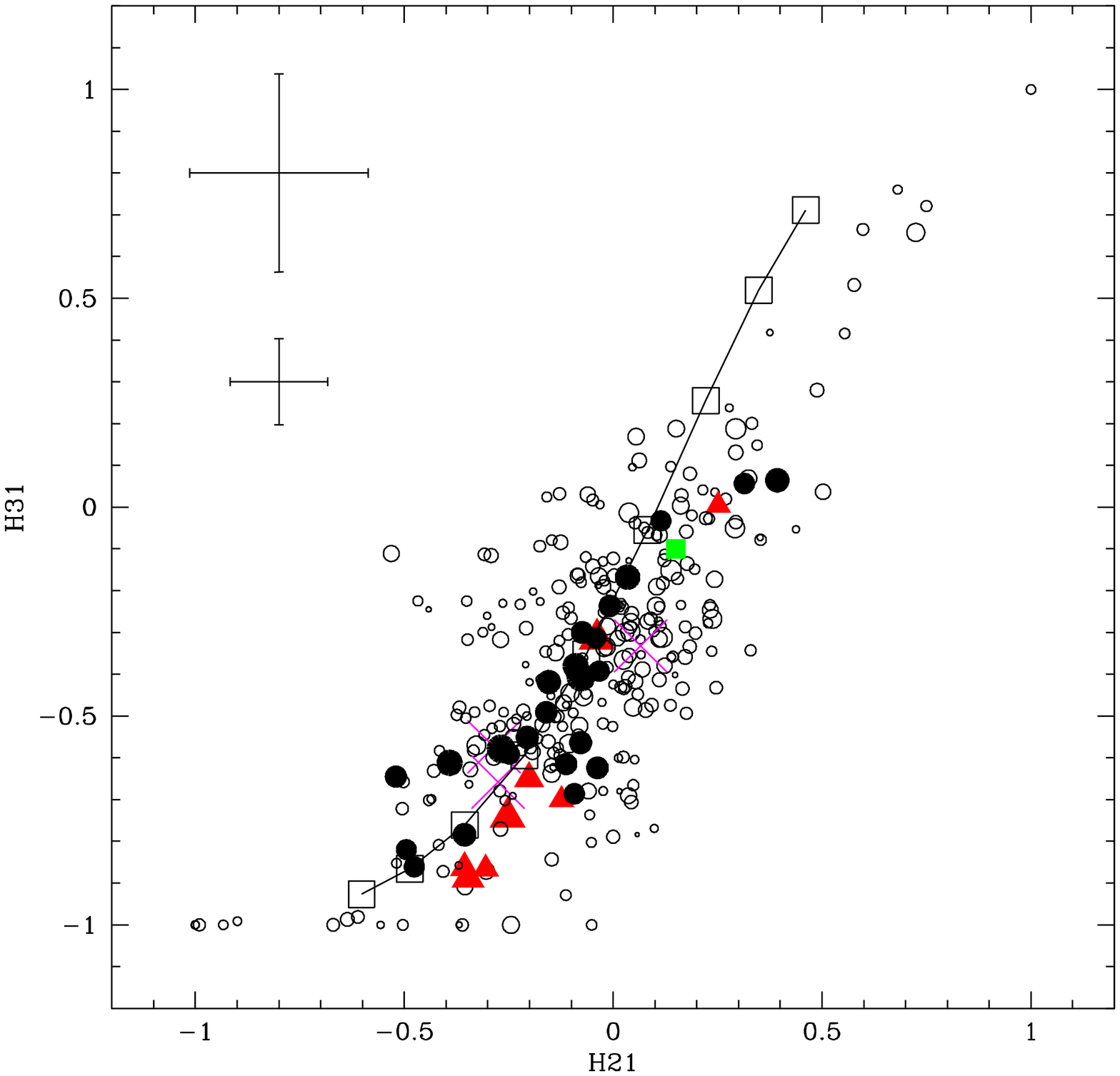}
\figcaption{\small
X-ray hardness ratios of point sources within NGC~1291, NGC~1316, NGC~1399,
NGC~1549, NGC~1553, NGC~4374, NGC~4472, NGC~4494, NGC46~36, NGC~4649,
and NGC~4697. The area of the circle is proportional
to the X-ray luminosity of the source. Filled circles represent sources
with luminosities of $1-2 \times 10^{39}$ ergs s$^{-1}$,
and open circle represent sources less luminous than $10^{39}$ ergs s$^{-1}$.
Red triangles represent sources that would have luminosities greater
than $2 \times 10^{39}$ ergs s$^{-1}$ if they were at the distance of
the galaxies, although they are most likely bright background/foreground
objects. Known AGN are symbolized by large magenta crosses. The colors predicted
from a typical ULX spectrum ($kT_{in}$ = 1.5 keV) is represented by a green
square. To eliminate scatter, only sources with at least 30 counts were
included. Error bars for a 50 count source and a 100 count source (typical of
$>10^{39}$ ergs s$^{-1}$ sources) are shown in the upper left.
The open squares represent the colors predicted from an absorbed power law model
with an exponent $\Gamma=0$ (upper right) to $\Gamma=3.2$ in increments of 0.4.
\label{fig:colors}}

\bigskip
Since the number of $>10^{39}$ ergs s$^{-1}$ sources is small, it is important
to check if there could be significant contamination from background AGN
at these flux levels, especially considering the apparent change in source
colors across the $10^{39}$ ergs s$^{-1}$ threshold. In
Table~\ref{tab:highflux}, we compare the number of sources in each galaxy
with fluxes corresponding to luminosities of
$1-2 \times 10^{39}$ ergs s$^{-1}$ and $>2 \times 10^{39}$ ergs s$^{-1}$
to the number of foreground/background sources expected over the area
covered, both for each galaxy and for all the galaxies combined.
We have calculated the expected number of foreground/background
sources
using the 0.5--2.0 keV log $N$--log $S$ relations of both 
Mushotzky et al. (2000; hereafter M00) and
Giacconi et al. (2001; hereafter G01) and extrapolating to a 0.5--6.0 keV
{\it Chandra} count flux using the spectral model assumed in each paper.
These two
{\it Chandra} deep field studies predict somewhat different numbers of high
flux sources, as illustrated in Figure~2 of Giacconi et al.\ (2001). This is
most likely the result of small number statistics of the number of sources
at the high flux end of the distribution. We have excluded NGC~1407 from
the combined results because of the uncertainty in its distance, which
greatly affects the number of high luminosity sources it contains (see
\S~\ref{subsec:classes} below). We have also excluded the two spiral
bulges, NGC~1291 and M31, for which no luminous sources were detected.
For sources with inferred luminosities between
$1-2 \times 10^{39}$ ergs s$^{-1}$, there are 4--5 times more sources than
expected from foreground/background sources, so clearly a
majority of these sources belong to the galaxies in our sample.
Conversely, the ten sources with inferred luminosities greater
than $2 \times 10^{39}$ ergs s$^{-1}$ are what is expected from
foreground/background sources. We note that two of the ten sources
are known to be contained within globular clusters of NGC~1399
(Angelini et al.\ 2001). The remaining eight sources are what is
expected if the M00 and G01 predictions are averaged. The colors of these
sources tend to be quite soft, and are denoted by red triangles in
Figure~\ref{fig:colors}. This is in accordance with the G01 finding that
the bright sources in their deep field had softer spectra than the
faint sources.

\begin{table*}[t]
\small
\caption{Number of Sources Detected vs.\ Number Expected From
Background AGN}
\label{tab:highflux}
\begin{center}
\begin{tabular}{lccccccc}
\multicolumn{8}{c}{} \cr
\tableline \tableline
&\multicolumn{3}{c}{$10^{39}$ ergs s$^{-1} < L_X <
2 \times 10^{39}$ ergs s$^{-1}$} &&
\multicolumn{3}{c}{$L_X > 2 \times 10^{39}$ ergs s$^{-1}$}  \\
\cline{2-4} \cline{6-8}
Galaxy &  Number & Number & Number && Number & Number & Number \\
& Expected & Expected & Detected && Expected & Expected & Detected \\
& M00 & G01 &&& M00 & G01 & \\
\tableline
NGC~1316 & 1.1 & 0.8 & 6 && 1.7 & 1.0 & 1  \\
NGC~1399 & 0.3 & 0.2 & 5 && 0.5 & 0.3 & 3  \\
NGC~1407\tablenotemark{a} & 1.5 & 1.3 & 7 && 2.3 & 1.6 & 5  \\
NGC~1407\tablenotemark{b} & 0.7 & 0.6 & 2 && 1.2 & 0.7 & 0  \\
NGC~1549 & 0.4 & 0.3 & 2 && 0.6 & 0.4 & 0  \\
NGC~1553 & 0.6 & 0.4 & 1 && 0.9 & 0.5 & 1  \\
NGC~3115 & 0.2 & 0.1 & 0 && 0.3 & 0.2 & 0  \\
NGC~3585 & 0.3 & 0.3 & 1 && 0.5 & 0.3 & 0  \\
NGC~4374 & 0.4 & 0.3 & 0 && 0.7 & 0.4 & 1  \\
NGC~4472 & 0.9 & 0.7 & 3 && 1.4 & 0.8 & 0  \\
NGC~4494 & 0.2 & 0.2 & 1 && 0.4 & 0.2 & 1  \\
NGC~4636 & 0.7 & 0.5 & 0 && 1.1 & 0.7 & 2  \\
NGC~4649 & 0.6 & 0.5 & 6 && 1.0 & 0.6 & 1  \\
NGC~4697 & 0.5 & 0.3 & 1 && 0.8 & 0.4 & 0  \\
\tableline
Total\tablenotemark{c} & 6.2 & 4.7 & 26 && 9.9 & 5.8 & 10\tablenotemark{d} \\
&&&&&&& \\
$0-1$ r$_{eff}$ & 1.0 & 0.7 & 10 && 1.5 & 0.9  & 1 \\
$1-3$ r$_{eff}$ & 5.2 & 4.0 & 16 && 8.4 & 4.9 & 9\tablenotemark{d} \\
\tableline
\end{tabular}
\end{center}
\tablenotetext{a}{Assuming a distance $d=28.8$ Mpc.}
\tablenotetext{b}{Assuming a distance $d=17.6$ Mpc.}
\tablenotetext{c}{Excluding NGC~1407.}
\tablenotetext{d}{Includes two sources known to reside in globular clusters
of NGC~1399 and two known AGN.}
\end{table*}

\subsection{Composite X-ray Spectrum for Sources with
$L_X < 10^{39}$ ergs s$^{-1}$}
\label{subsec:composite}

Given the tendency of the most luminous sources to be softer than the lower
luminosity sources, we have re-determined the best-fit power law exponent
for the spectrum of all the sources in each galaxy excluding the sources
whose luminosities exceeded $10^{39}$ ergs s$^{-1}$ as well as the
supersoft sources. The results are shown in Figure~\ref{fig:combine}. Upon
removal of the most luminous sources, the spectra of the remaining sources
appear remarkably uniform from galaxy to galaxy. The unweighted mean and
standard deviation of the sample was $\Gamma = 1.56 \pm 0.08$. We performed a
simultaneous fit to the 15 data sets within XSPEC, allowing the normalizations
for each galaxy to vary while having one common power law exponent.
The best-fit exponent was $\Gamma = 1.56\pm0.02$ (90\% confidence level),
with $\chi_{\nu}^2 = 1.08$ for 903 degrees of freedom. A thermal
bremsstrahlung model led to a slightly worse fit ($\chi_{\nu}^2$ of 1.14)
with $kT_{brem}=7.3 \pm 0.3$ keV. A disk blackbody model led to a
statistically unacceptable fit ($\chi_{\nu}^2$ = 1.79).

\subsection{Dependence of LMXB Spectral Properties on Source Luminosity} 
\label{subsec:classes}

We further subdivided the sources into four luminosity classes for each
galaxy: (1) $L_X = 1-2 \times 10^{39}$ ergs s$^{-1}$,
(2) $10^{38}$ ergs s$^{-1} < L_X < 10^{39}$ ergs s$^{-1}$,
(3) $10^{37}$ ergs s$^{-1} < L_X < 10^{38}$ ergs s$^{-1}$, and
(4) $L_X < 10^{37}$ ergs s$^{-1}$. Only the
bulge of M31 is close enough to contribute to the final class.
For the most luminous class, only NGC~1316, NGC~1399, and NGC~4649
are used, since these are the only galaxies where there is clearly an
excess of sources over that expected from foreground/background sources.
Figure~\ref{fig:classes} shows the best-fit power law exponents for
the three lower luminosity classes for each galaxy. There was
no statistical difference between classes (2) and (3).
\medskip
\centerline{\null}
\vskip3.30truein
\includegraphics{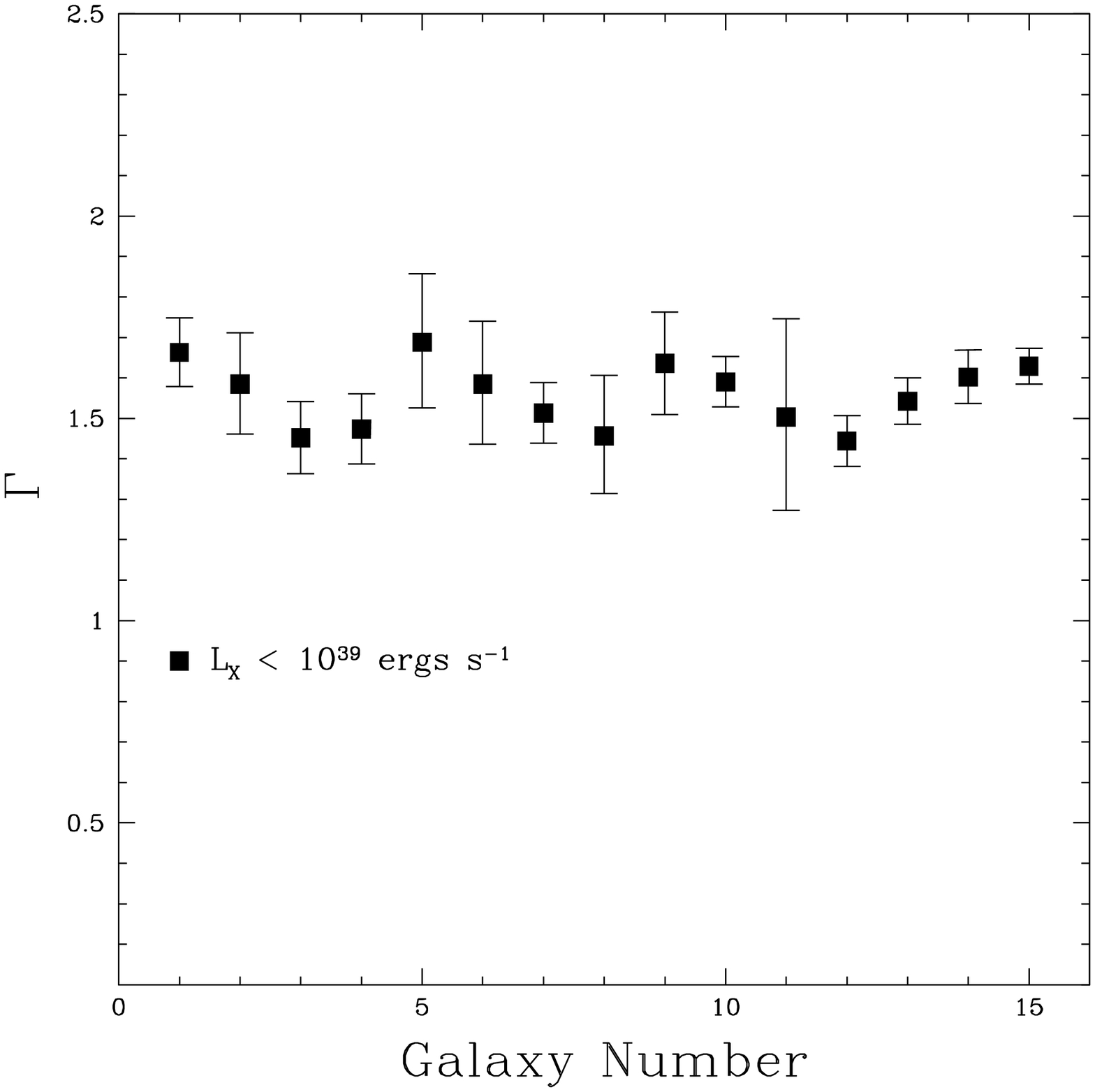}
\figcaption{\small
Best-fit power law exponent ($\Gamma$) when the spectra of the sources were
summed for each galaxy in our sample. The galaxy number corresponds to
the numbering system in Table~\protect\ref{tab:sample}.
\label{fig:combine}}
\bigskip
\noindent Interestingly, the spectrum of
the lowest luminosity source
of M31) is indistinguishable from sources in the $10^{37} - 10^{39}$
ergs s$^{-1}$ regime. Thus, it appears on average that the bulk spectral
characteristics of LMXBs are the same over at
\bigskip
\centerline{\null}
\vskip3.30truein
\includegraphics{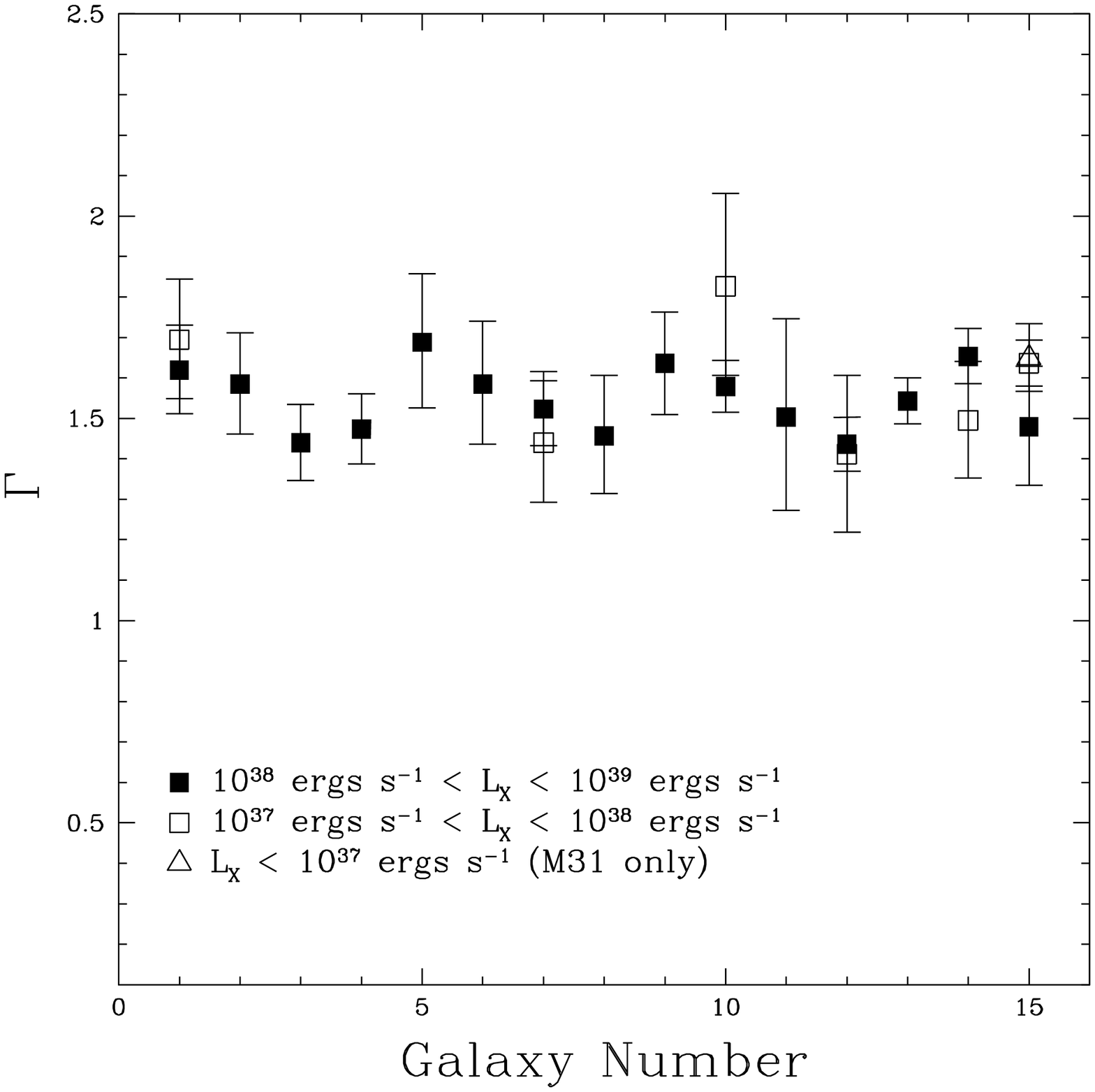}
\figcaption{\small
Best-fit power law exponents for each galaxy for three luminosity
classes below $10^{39}$ ergs s$^{-1}$.
\label{fig:classes}}
\noindent least three orders of magnitudes
in luminosity up to $10^{39}$ ergs s$^{-1}$.
\begin{table*}[b]
\small
\caption{Summary of Power Law Spectral Properties For Each Galaxy By
Luminosity Class}
\label{tab:classes}
\begin{center}
\begin{tabular}{lccccccc}
\multicolumn{8}{c}{} \cr
\tableline \tableline
Number & Galaxy & All sources & $\chi_{\nu}^2$/d.o.f &
$< 10^{39}$ ergs s$^{-1}$ &$\chi_{\nu}^2$/d.o.f &
1--2 $\times 10^{39}$ ergs s$^{-1}$ &$\chi_{\nu}^2$/d.o.f  \cr
&& ($\Gamma$) && ($\Gamma$) && ($\Gamma$) & \cr
\tableline
1 & NGC~1291 & $1.66^{+0.09}_{-0.08}$& 1.27/49 & $1.66^{+0.09}_{-0.08}$& 1.27/49
& \ldots & \ldots \\
2 & NGC~1316 & $1.86^{+0.12}_{-0.11}$& 0.71/41& $1.59^{+0.13}_{-0.12}$&0.69/26
& $1.86^{+0.28}_{-0.26}$& 1.58/13 \\
3 & NGC~1399 & $1.72^{+0.06}_{-0.06}$& 0.94/126& $1.45^{+0.09}_{-0.09}$&1.08/58
& $1.95^{+0.14}_{-0.14}$& 0.66/25 \\
4 & NGC~1407 & $1.56^{+0.06}_{-0.06}$& 1.41/85& $1.47^{+0.09}_{-0.09}$&1.08/58
& $1.37^{+0.19}_{-0.19}$&0.97/14 \\
5 & NGC~1549 & $1.61^{+0.16}_{-0.16}$& 0.56/17& $1.69^{+0.17}_{-0.16}$&0.65/13
& \ldots & \ldots \\
6 & NGC~1553 & $1.69^{+0.11}_{-0.11}$& 1.33/31& $1.59^{+0.15}_{-0.15}$&1.28/23
& \ldots & \ldots \\
7 & NGC~3115 & $1.59^{+0.07}_{-0.07}$& 1.06/61& $1.51^{+0.08}_{-0.07}$&1.07/56
& \ldots & \ldots \\
8 & NGC~3585 & $1.43^{+0.15}_{-0.15}$& 1.32/18& $1.46^{+0.15}_{-0.15}$&1.87/16
& \dots &\ldots \\
9 & NGC~4374 & $1.89^{+0.08}_{-0.08}$& 0.84/66& $1.64^{+0.12}_{-0.13}$&0.83/32
& \ldots & \ldots \\
10 & NGC~4472 & $1.57^{+0.06}_{-0.06}$& 0.98/116&$1.59^{+0.06}_{-0.06}$&0.96/107
& \dots &\ldots \\
11 & NGC~4494 & $1.87^{+0.17}_{-0.16}$& 0.82/19 &$1.50^{+0.25}_{-0.23}$&0.86/8
&  \ldots & \ldots \\
12 & NGC~4636 & $1.57^{+0.05}_{-0.05}$& 0.97/105&$1.44^{+0.06}_{-0.06}$&0.84/88
& \ldots & \ldots \\
13 & NGC~4649 & $1.61^{+0.04}_{-0.04}$& 1.17/122&$1.54^{+0.06}_{-0.06}$&1.30/105& $1.77^{+0.12}_{-0.12}$& 1.26/26 \\
14 & NGC~4697 & $1.63^{+0.06}_{-0.06}$& 1.40/80& $1.60^{+0.07}_{-0.07}$&1.23/73
& \ldots & \ldots \\
15 & M31 &  $1.63^{+0.04}_{-0.04}$& 1.18/133& $1.63^{+0.04}_{-0.04}$& 1.18/133
& \ldots & \ldots \\

\tableline
\end{tabular}
\end{center}
\end{table*}
\begin{table*}[t]
\caption{Summary of Simultaneous Power Law Spectral Fits}
\label{tab:simultaneous}
\begin{center}
\begin{tabular}{lcccc}
\multicolumn{5}{c}{}  \cr \cr
\tableline \tableline
Category && $\Gamma$ &  & $\chi_{\nu}^2$/d.o.f  \cr
\tableline
All Sources, All Galaxies  &  &$1.64^{+0.02}_{-0.02}$&& 1.15/1083 \\
$L_X = 1-2 \times 10^{39}$ ergs s$^{-1}$  & &$1.96^{+0.08}_{-0.08}$&& 1.15/70 \\
$L_X < 10^{39}$ ergs s$^{-1}$  &  &$1.56^{+0.02}_{-0.02}$&& 1.08/903 \\
$10^{38}$ ergs s$^{-1} < L_X <  10^{39}$ ergs s$^{-1}$
&  &$1.56^{+0.02}_{-0.03}$&& 1.04/727 \\
$10^{37}$ ergs s$^{-1} < L_X < 10^{38}$ ergs s$^{-1}$
&  &$1.61^{+0.05}_{-0.05}$&& 1.10/189 \\
$L_X < 10^{37}$ ergs s$^{-1}$ (M31 bulge only )&  &$1.64^{+0.06}_{-0.06}$&& 1.32/92 \\
Sources $< 1r_{eff} $ &  &$1.48^{+0.04}_{-0.04}$&& 1.13/320 \\
Sources $1r_{eff}-2r_{eff}$ &  &$1.51^{+0.03}_{-0.04}$&& 0.94/299 \\
Sources $2r_{eff}-3r_{eff}$ &  &$1.53^{+0.06}_{-0.06}$&& 1.03/134 \\

\tableline
\end{tabular}
\end{center}
\end{table*}

The situation is quite different for sources with luminosities of
$1-2 \times 10^{39}$ ergs s$^{-1}$ as alluded to earlier.
Since several galaxies contained only one or
two such high flux sources, a single contaminating background AGN could
alter the best-fit power law index significantly if misidentified as an
X-ray source within the target galaxy. We have checked to see if any
of the luminous sources coincided spatially with known AGN. The only high flux
source in both NGC~4374 and NGC~4697 are known AGN and have been excluded.
NGC~4636 and NGC~4649 each have a
bright source that is coincident with optical identifications from
the USNO-A2.0 optical catalog (Monet et al. 1998).
The $B-R$ value given in the USNO-A2.0 optical for the
source in NGC~4649 is --0.7, far too blue to be a globular cluster of
NGC~4649 and is most likely an AGN. The bright source in NGC~4636 has
$B-R= 1.7$, which would make it red enough to be a globular cluster. However,
such an identification is premature,
since a known AGN near NGC~4374 has a similar red color. Given the 
uncertainty in determining which high luminosity sources were actually
AGN, we have calculated power law indices for
only galaxies that contained at least four such sources
(see Table~\ref{tab:classes}).
NGC~1316, NGC~1399, and NGC~4649 all have power law indices
significantly higher than the lower luminosity sources, as the
colors information of Figure~\ref{fig:colors} implied.
NGC~1407 has a low power law index ($\Gamma = 1.37 \pm 0.19$) despite
containing seven high luminosity sources,
but there is considerable uncertainty in the distance to this galaxy.
While we have assumed the Tonry et al.\ (2001) surface brightness
fluctuation distance of 28.8 Mpc, the globular cluster luminosity function
distance method gives a distance of only 17.6 Mpc (Perrett et al.\ 1997).
With this smaller distance, only two of the sources would have a luminosity
exceeding $10^{39}$ ergs s$^{-1}$, with $\Gamma = 1.75\pm0.21$, which is
more similar to the values obtained for NGC~1316, NGC~1399, and NGC~4649.
Also using this distance, the seven sources that yielded
$\Gamma = 1.37 \pm 0.19$ would all have luminosities less than
$10^{39}$ ergs s$^{-1}$, and the low value of $\Gamma$ would be consistent
with that expected from lower luminosity sources in other galaxies.
The spectra of the LMXBs in NGC~1407 would at least tentatively argue
in favor of the lower distance estimate to this galaxy.

If we make a simultaneous fit for all the galaxies
that have at least four high luminosity sources (and also exclude NGC~1407
because of the uncertainty in its distance)
so that the effects of including an unidentified AGN are minimized,
we find a best-fit power law exponent of $\Gamma = 1.96 \pm 0.08$.
A summary of the best-fit power laws by luminosity class above and below
$10^{39}$ ergs s$^{-1}$ for each galaxy is
given in Table~\ref{tab:classes}, and a summary of the simultaneous
fits by luminosity class is given in Table~\ref{tab:simultaneous}.

Although it has been postulated that the break in the luminosity function
at $\sim3 \times 10^{38}$ ergs s$^{-1}$ represents a transition between neutron
star and black hole LMXBs, there is no apparent difference in spectral
characteristics of sources above and below the break (excluding the
very luminous $>10^{39}$ ergs s$^{-1}$ sources). Fitting the spectra of
all sources in all the galaxies with luminosities less than
$3 \times 10^{38}$ ergs s$^{-1}$ yielded a best-fit power law spectrum with
$\Gamma = 1.58\pm0.03$, while fitting sources more luminous than 
$3 \times 10^{38}$ ergs s$^{-1}$ (but less than $10^{39}$ ergs s$^{-1}$)
yielded $\Gamma = 1.54\pm0.03$. This is expected
given that differences in the spectral signatures between
black holes in their non-flaring state and neutron stars differ appreciably
only at hard X-ray energies ($\ga$ 5 keV). A disk blackbody model
provided a poor fit ($\chi_{\nu}^2 > 1.5$) for both luminosity classes.

\subsection{Globular Cluster vs.\ Non-globular Cluster Sources}
\label{subsec:globular}

We have tested whether the spectral properties of LMXBs that
reside within globular clusters differ from those not residing within globular
clusters. For NGC~1399 and NGC~4472, we used the globular cluster--X-ray
source identifications of Angelini et al.\ (2001) and Kundu et al.\
(2002), respectively, to derive best-fit power laws for LMXBs residing
inside and outside of globular clusters. We excluded sources with luminosities
that exceeded $10^{39}$ ergs s$^{-1}$. For NGC~1399, globular cluster
LMXBs had a best-fit power law index of $\Gamma = 1.33\pm0.17$, while for
non-globular cluster LMXBs, $\Gamma = 1.62\pm0.34$. A similar exercise for
NGC~4472 found $\Gamma = 1.51\pm0.13$ for globular cluster sources and
$\Gamma = 1.57\pm0.06$ for non-globular cluster sources. A similar result
was found for this galaxy by Maccarone, Kundu, \& Zepf (2002). So within the
uncertainties, there is no difference in the spectral properties of LMXBs
based on their location inside or outside a globular cluster.

\subsection{Radial Dependence of the Spectral Properties of LMXBs}
\label{subsec:radial}

We also investigated the radial dependence of the spectral properties of
LMXBs. Sources were extracted from within 0--1, 1--2, and 2--3 effective
radii (again the highest luminosity and supersoft sources were excluded)
and a best-fit power law exponent was obtained. For each radial bin, the
spectra from all the galaxies were fit simultaneously as described above
(the spiral bulges of NGC~1291 and M31 were excluded from the fit).
For the three radial bins, the best-fit exponents were $\Gamma = 1.48\pm0.05$,
$\Gamma = 1.51\pm0.05$, and $\Gamma = 1.62\pm0.06$ (90\% uncertainties),
respectively. Although
the first two radial bins are consistent with one another, the first and
third bins are marginally inconsistent. We investigated if this was due to
contamination from unrelated foreground/background sources in the third
radial bin. For each galaxy, we used the {\it Chandra} deep field source
count rate of M00 to estimate the expected number of serendipitous
sources in the third radial bin. If the number of expected contaminating sources
exceeded 30\% of the total number of sources actually detected in this radial
bin, we excluded it from the simultaneous fit. After excluding NGC~1316,
NGC~1553, NGC~4472, NGC~4494, and NGC~4697 because of this criterion, the
best-fit
power law exponent from a simultaneous fit of the remaining eight galaxies
was $\Gamma = 1.53\pm0.06$, which lessens the difference with the inner
two radial bins.
Thus, there appears to be no radial dependence of the spectral properties
of LMXBs at least out to three effective radii. This information is summarized
in Table~\ref{tab:simultaneous}.

Given that five of the galaxies had significant contamination in the
third radial bin, we re-determined the best-fit power law exponent
for all sources with $L_X < 10^{39}$ ergs s$^{-1}$ sources without
including sources in the third radial bin. However, the results before
and after the exclusion were consistent within the uncertainties. Also,
the composite spectrum was not changed significantly.

\section{Discussion}
\label{sec:discussion}

\subsection{Comparison with Previous Studies}
\label{subsec:comparison}

The constraints on the bulk spectral properties of LMXBs of
$\Gamma = 1.56\pm0.02$ (or alternatively $kT_{brem}=7.3 \pm 0.3$ keV)
represent the tightest constraints derived to date. Previously,
Matsumoto et al.\ (1997) simultaneously fit the spectra of several early-type
galaxies observed with {\it ASCA} with a two component model to fit jointly
the hot gas and LMXB component; the best-fit power law model yielded
$\Gamma = 1.8 \pm 0.4$. More recently, White (2001)
also used {\it ASCA} data for six early-type galaxies
to set a constraint of $\Gamma = 1.83^{+0.10}_{-0.11}$ (90\% confidence
limits) on the power law index, in marginal disagreement with our result
(although their best-fit bremsstrahlung temperature of
$kT_{brem}= 6.3^{+1.6}_{-1.1}$ is consistent with our value).
However, since the rather poor spatial resolution of {\it ASCA} could not
in general resolve most of the sources, the more luminous sources
undoubtedly biased the power law exponent higher, owing to the soft spectral
nature of these luminous sources. If we make a composite fit to the spectra
from all 15 galaxies without excluding the most luminous sources, we obtain
$\Gamma = 1.64\pm0.02$, in closer agreement with the White (2001)
result. Furthermore, three of the galaxies in our sample have
$\Gamma \sim 1.9$ (Figure~\ref{fig:all}). If these three galaxies
were part of a smaller sample than what we have compiled here, we would
have obtained an average value closer to $\Gamma = 1.8$ for the sample.

\subsection{High/Soft vs.\ Low/Hard Spectral States?}
\label{subsec:hard_soft}

The result that sources with luminosities of $1-2 \times10^{39}$ ergs s$^{-1}$
are substantially softer on average than sources less luminous than
$10^{39}$ ergs s$^{-1}$ is highly reminiscent of the high flux/soft spectral
state vs.\ low flux/hard spectral shape behavior exhibited by Galactic
black hole X-ray binaries candidates (e.g., Tanaka \& Lewin 1995; Nowak 1995).
While in their low/hard state, the spectra of Galactic black hole X-ray binaries
are characterized by a power law with $\Gamma = 1.3-1.7$ with an exponential
cut-off at about 100 keV, while in their high/soft state, they exhibit a
disk blackbody component ($kT_{in} \sim 1$ keV) plus a power law component
with $\Gamma \sim2.5$. While the high energy cut-off of the low/hard state is
obviously impossible to detect in our sample, our best-fit power law of
$\Gamma = 1.56\pm0.02$ for sources with $L_X < 10^{39}$ ergs s$^{-1}$ is
certainly consistent with Galactic examples in their low/hard state.
Intriguingly, if we fit simultaneously the most luminous sources in our sample
with a disk blackbody plus power law model, we get best-fit
values of $kT_{in} = 1.76^{+0.16}_{-0.23}$ keV and
$\Gamma = 2.61^{+0.51}_{-0.40}$, with the power law component contributing
between 25\%--68\% of the total 1--10 keV luminosity from galaxy to galaxy.
This ratio is that expected from black hole binaries in the
very high state of Figure 1 of Nowak (1995), in which the binaries
are accreting close to the Eddington limit. 
This spectral model was also a better fit than the single power law
model ($\chi_{\nu}^2$ = 0.89/66 d.o.f. vs.\ $\chi_{\nu}^2$ = 1.15/70 d.o.f.).

Finally, it is unlikely that the luminous sources found in elliptical and S0
galaxies are actually multiple lower-luminosity sources.
If this were the case,
one would not expect the difference in spectral characteristics between these
sources and lower luminosity sources. A single accreting object provides the
most reasonable explanation.

\subsection{Spatial Distribution of the Luminous Sources}
\label{subsec:spatial}

The spatial distribution of the luminous sources can yield useful insight
into the nature of these sources. We have determined how many
$>10^{39}$ ergs s$^{-1}$ sources are found
between 0--1 $r_{eff}$ and 1--3 $r_{eff}$, respectively, for all
13 elliptical and S0 galaxies in our sample collectively. Since these sources
typically have 100 or more X-ray counts, the effects of missing sources because
of the degradation of the point spread function with increasing off-axis
distance or from contamination from high X-ray surface brightness hot gas
in some of the galaxies is minimized and can be safely ignored. 

If the sources are distributed randomly over the field of view of the
ACIS detector, we would expect eight times more sources in the
1--3 $r_{eff}$ bin than in the 0--1 $r_{eff}$ bin. Because
the 3 $r_{eff}$ radius contour did not always fit onto the ACIS chip for the
larger galaxies, the ratio for our sample is expected to be about 5.5:1, rather
than 8:1. Conversely, if the sources follow the optical light, one would
expect about half as many sources within the 1--3 $r_{eff}$ bin than in
the 0--1 $r_{eff}$ bin.

For the $> 2 \times 10^{39}$ ergs s$^{-1}$ sources, once the two sources
known to be associated with globular clusters of NGC~1399 are removed, there
are seven sources in the 1--3 $r_{eff}$ bin and only one within
one effective radius (Table~\ref{tab:highflux}). This is consistent with
the sources being randomly distributed over the detector, providing further
evidence that these sources are unrelated foreground/background sources.
The spatial distribution of the $1-2 \times 10^{39}$ ergs s$^{-1}$ sources
is clearly not random. Once an estimate for the foreground/background
sources is removed, there are about nine sources within one effective
radius and 11 sources within 1--3 $r_{eff}$. Thus, the spatial
distribution of these sources is much flatter than the optical light, in
which there should have been twice the number of sources in the inner
spatial bin than in the outer one. This is not unexpected, given that
a large percentage of the sources are within globular clusters, and that
the spatial distribution of globular cluster within early-type galaxies is
known to be significantly flatter than the optical light, which follows
a de Vaucouleurs profile. In fact, the two high luminosity sources within
both NGC~1399 and NGC~4472 that fall within the
{\it Chandra}--{\it Hubble Space Telescope} overlap region of the galaxies
reside within globular cluster. Further optical work will be required
to determine if all the high luminosity sources reside within globular
clusters, but the extended spatial distribution of the sources would
suggest that they do.

\subsection{Are The Luminous Sources Ultraluminous X-ray Sources (ULXs)?}
\label{subsec:ulxs}

The existence of very luminous off-center X-ray point sources in spiral
galaxies has been the subject of considerable interest in recent years
(e.g., Makishima et al.\ 2000; Roberts \& Warwick 2000; King et al.\ 2001).
Usually found in star-forming regions of spiral arms and having
$L_X > 10^{39}$ ergs s$^{-1}$ (although this lower limit on the luminosity
is somewhat arbitrary), the nature
of these ultraluminous X-ray sources has been difficult to explain.
Although a few ULXs appear to be supernovae that detonated in a very
dense environment (Fabian \& Terlevich 1996; Immler et al.\ 1998;
Blair, Fesen, \& Schlegel 2001), significant
temporal variability has ruled out this possibility in many other cases,
strongly indicating that they are some form of accreting X-ray binary.
However, ULXs in NGC~1313, M81, and IC~342 do not exhibit the typical
high/soft--low/hard flux/spectral behavior that black hole binaries exhibit
but instead show a reverse behavior (Makishima et al.\ 2000;
Mizuno, Kubota, \& Makishima 2001).
Still, the fact that ULX spectra can be adequately fit with a disk
blackbody model argues that they are an accreting binary of some kind.

The high X-ray luminosities of the more luminous ULXs imply that the mass
of the central accreting object must exceed 50 M$_{\odot}$ if they are
accreting at the Eddington limit (where $L_{Eddington} = 1.3 \times 10^{38}$
ergs s$^{-1}$ $M$/M$_{\odot}$ and $M$ is the mass of the accreting object),
assuming the emission is unbeamed. Sub-Eddington accretion rates typical
of Galactic X-ray binaries would imply an even
higher black hole mass. Such a class of intermediate mass black holes would
provide the crucial missing link between stellar mass ($\la 10$ M$_{\odot}$)
black holes and supermassive ($> 10^6$ M$_{\odot}$) black holes at the centers
of galaxies. However, black holes of this size are very difficult to create
from the collapse of single stars, given that current stellar evolution
models predict that massive main-sequence progenitor stars will lose too
much mass via stellar winds throughout the course of its lifetime to retain
a $>50$ M$_{\odot}$ core to produce such a massive black hole. This has created
some skepticism of the intermediate mass black hole explanation. Alternatively,
the X-ray emission might not be isotropic but instead beamed at us.
King et al.\ (2001) has proposed that the most likely candidate in a
beaming scenario is a period of thermal time scale mass transfer in
binaries with intermediate or high mass donor stars, which naturally explains
the presence of the ULXs in regions of recent star formation.

While the focus of ULX studies has been on sources in spiral galaxies,
information gathered on possible ULXs in elliptical galaxies has yet to be
incorporated into their explanation. A substantial number of high luminosity
X-ray sources in
early-type galaxies was found by Colbert \& Ptak (2002) in a survey with the
{\it ROSAT} HRI, although the lack of spectral resolution of that instrument
did not allow for a spectral comparison of these sources to ULXs in
spiral galaxies.

From our {\it Chandra} sample, we find that very luminous ($> 2 \times
10^{39}$ ergs s$^{-1}$) sources are absent from elliptical and S0 galaxies,
save for two luminous sources within globular clusters of NGC~1399.
Furthermore, the sources in our sample with luminosities of
$1-2 \times 10^{39}$ ergs s$^{-1}$ have different spectra than ULXs in
spiral galaxies. A single disk blackbody model does not provide a good
fit to the composite spectrum of the high luminosity sources within NGC~1316,
NGC~1399, and NGC~4649 ($\chi_{\nu}^2$ of 2.74 for 70 degrees of freedom)
unlike spiral galaxy ULXs (e.g., Makishima et al.\ 2000).
The addition of a power law tail is required at very high significance,
something required for only one ULX studied to date (M33 X-8;
Makishima et al.\ 2000).
The disk blackbody + power law model required to fit the high
luminosity sources (\S~\ref{subsec:hard_soft}) is much more similar
to the model used to fit high state
black hole binaries than the model used to fit ULXs. This is illustrated
in Figure~\ref{fig:colors}, where the colors predicted from a disk
blackbody with in inner temperature of $kT_{in} = 1.5$ keV (green square)
are significantly different that the colors of the high luminosity sources.

The two very luminous ($\sim 4 \times 10^{39}$ ergs s$^{-1}$) sources
found within globular clusters of NGC~1399 by Angelini et al.\ (2001)
appear to be very rare objects indeed. It is unlikely that any other such
high luminosity sources exist in our sample given that the number of
expected serendipitous foreground/background objects should account for
all the other high flux sources in our sample, statistically speaking,
although further optical work will be needed to search for possible optical
counterparts of the very luminous sources.
If such rare, very luminous sources are specific to globular clusters
then perhaps it is not surprising that both are found within globular
clusters of NGC~1399 given that NGC~1399 has such a large number of
globular clusters (it has a globular cluster specific frequency that is
2--3 times that of typical elliptical galaxies; Harris 1991).

The combined spectra of the two globular cluster sources also look more
like that of a high state black hole binary than a ULX. A disk blackbody
alone provided a very poor fit to the data
($\chi_{\nu}^2$ of 2.48 for 32 degrees of freedom), while a simple power
law provided a very good fit with $\Gamma = 1.93 \pm 0.11$, very similar
to the best-fit power law exponent for the $1-2 \times 10^{39}$ ergs s$^{-1}$
sources, although in this instance the addition of a disk blackbody model
to the power law did not significantly improve the fit. Thus, it is
possible that these very luminous sources are not ULXs (although they have
ULX-like luminosities), and represent a different type of accreting object
than that responsible for ULXs in spiral galaxies.

One possibility for these very luminous globular clusters sources is
accretion onto a central intermediate mass black hole at the center of
the globular cluster. The formation of luminous X-ray sources
within globular clusters has been discussed in detail by
Miller \& Hamilton (2002). In their scenario, a $\ga 50$ M$_{\odot}$ black hole
sinks to the center of the globular cluster and accretes smaller black holes
over the lifetime of the cluster to grow to a mass of $\sim1000$ M$_{\odot}$.
The creation of the initial
$50$ M$_{\odot}$ black hole (the smallest mass black hole that would not
be ejected from the cluster by recoil during three-body interactions)
might not be a problem in the low metallicity
environment of a globular cluster, since mass-loss rates of very massive stars
is predicted to be mild to negligible for low metallicity stars (see, e.g.,
Vink, de Koter \& Lamers 2001). If the accretion efficiency of such a black
hole was 0.01--0.1 as in Galactic black holes, X-ray luminosities of
$10^{39}-10^{40}$ ergs s$^{-1}$ could be achieved. Such a scenario alleviates
the problem of having the black hole emit persistently near its Eddington
limit.

If the well-known relation in elliptical galaxies and spiral bulges
between the stellar velocity dispersion and mass of the central black hole
(e.g., Gebhardt et al.\ 2000a) is extrapolated down to objects the size
of globular clusters, the relation predicts globular clusters will contain
central black holes with masses on the order of $10^3$ M$_{\odot}$. In fact,
kinematical evidence for a 2500 M$_{\odot}$ black hole in the center of the
Galactic globular cluster M15 (Gebhardt et al.\ 2000b) indicates that globular
clusters are indeed capable of harboring intermediate-mass black holes. Although
the lack of a very luminous X-ray source at the center of M15 indicates that
the black hole is not currently being fed, the presence of very luminous
X-ray sources in globular clusters might serve as an indication that
central massive black holes are a common feature of stellar systems over
a wide range of masses ranging from giant galaxies to globular clusters.
Demonstrating that the ULX is at the dynamical center of the globular
cluster, however, is not currently feasible even for the nearest early-type
galaxies.

In summary, it does not appear that elliptical galaxies contain ULXs,
at least not the kind found within spiral galaxies. For sources below
$2 \times 10^{39}$ ergs s$^{-1}$, the implied masses assuming the sources
are accreting near their Eddington limit are only $\la 15$ M$_{\odot}$,
which does not contradict the current estimates of the upper mass limit of
stellar mass black holes. Furthermore, their X-ray spectra are substantially
different than those of ULXs in spiral galaxies, as Figure~\ref{fig:colors}
indicates. Their spectral similarities to known black hole binaries strongly
argues that these sources are simply black holes accreting near their
Eddington limit, and do not require either beaming or an intermediate
mass black hole to explain their existence.
In addition, the only two sources with luminosities exceeding
$2 \times 10^{39}$ ergs s$^{-1}$ that are conclusively in the galaxies of
our sample (and probably the only two) also have much different spectra
than ULXs. The increasing evidence that globular clusters can contain
intermediate mass black holes also provides a method
for explaining their high X-ray luminosities that is problematic for spiral
galaxy ULXs that occur in the field.

\subsection{Implications for the Hot Gas Component}
\label{subsec:hot_gas}

The uniformity of the spectral properties of LMXBs with luminosities
below $10^{39}$ ergs s$^{-1}$ has useful implications for the
study of hot gas in early-type galaxies. First, in galaxies
where most of the LMXB emission is not resolved (either galaxies
observed with {\it XMM-Newton} or more distant galaxies where excessively
long {\it Chandra} observations would be needed to detect sources fainter
than $\sim 10^{38}$ ergs s$^{-1}$), the LMXB component will contribute
significantly to the diffuse emission, particularly for galaxies that
are not rich in hot gas. In this case, the ability to fix the spectral
shape of the LMXB component ($\Gamma = 1.56$ or $kT_{brem}=7.3$ keV) will
allow tighter constraints to be placed on the temperature and especially
the metallicity of the hot gas. All that is required is that LMXBs
more luminous than $10^{39}$ ergs s$^{-1}$ be removed from the spectrum.
As an example of this, we have extracted the spectrum of the diffuse
emission from within one effective radius of the X-ray faint galaxy NGC~3115,
and fit the emission with a MEKAL component
(Mewe, Gronenschild, \& van den Oord 1985; Kaastra \& Mewe 1993;
Liedahl et al.\ 1995) to model the hot gas component and a power law to
model the unresolved LMXB component. If the power law exponent is left
as a free parameter, a gas temperature of $kT_{MEKAL} = 0.61^{+0.28}_{-0.31}$
is obtained, and the metallicity of the gas is unconstrained. After fixing
the power law component at $\Gamma=1.56$, the temperature is constrained
significantly better, with $kT_{MEKAL} = 0.61^{+0.16}_{-0.18}$. In addition,
the metallicity of the gas could now be constrained to be less than
5\% of the solar value.

Secondly, the result that LMXBs in the bulge of M31 with luminosities
as low as $10^{36}$ ergs s$^{-1}$ have bulk spectral characteristics
that are indistinguishable from more luminous LMXBs will greatly simplify
the determination of the total gaseous X-ray luminosity, $L_{X,gas}$ in very
X-ray faint galaxies. In these types of galaxies, where very low X-ray count
rates will not allow the gas and LMXB components to be separated spectrally
as in NGC~3115, $L_{X,gas}$ can be very difficult to determine accurately.
A small error in the estimation of the unresolved LMXB component would lead to
a large error in the derived luminosity of the gaseous component. Although
such an error would have a negligible effect on the determination
of $L_{X,gas}$ in well-known, gas-rich galaxies such as NGC~4636 and
NGC~1399 (where the gaseous emission is over an order of magnitude more
luminous than the LMXB component), this will not be the case for a
galaxy such as NGC~3585. If we assume that all the diffuse counts in NGC~3585
above 2 keV are from unresolved LMXBs (since the emission from 0.3 keV gas
typical of such X-ray faint galaxies would be negligible above 2 keV),
we can convert the 2--6 keV flux to the 0.3--6 keV flux that
emanates from unresolved sources, and subtract this amount from the total
diffuse amount to yield the gaseous flux. If we use a $\Gamma = 1.56$ power
law model in the conversion, we find that 60\% of the 0.3--6 keV diffuse flux
is from
unresolved LMXBs, leaving 40\% for the hot gas component. Conversely, if we
assume a $\Gamma = 1.9$ power law model, LMXBs account for 83\% of the diffuse
emission, leaving only 17\% of the diffuse emission for the gas component.
Clearly, the derivation of
$L_{X,gas}$ is highly dependent on our choice of $\Gamma$ for the unresolved
sources, and factors of two or more uncertainties can arise. Knowing that we can
safely use a value of $\Gamma = 1.56$ will significantly decrease the
uncertainty in $L_{X,gas}$ for these very X-ray faint systems.

Determining
$L_{X,gas}$ accurately for X-ray faint galaxies is particularly important
in deriving the $L_{X,gas}$ vs.\ $L_{opt}$ relation, which has been used
extensively to characterize the hydrodynamical history of gas lost from
stars within the galaxy over the lifetime of the galaxy
(Canizares, Fabbiano, \& Trinchieri 1987; Davis \& White 1996;
Brown \& Bregman 1998). Comparison of the X-ray to optical luminosities of
galaxies found
$L_{X,gas} \propto L_B^{1.7-3.0}$ among the various studies.
The rather large spread in the best-fit exponent is in large part due to
uncertainties in the luminosity of the lower X-ray luminosity systems,
and how each investigator chose to subtract off the LMXB component.
Future studies with {\it Chandra} will minimize the uncertainty in the
$L_{X,gas}$ values and lead to a more accurate determination of the exponent.

\section{Conclusions}
\label{sec:conclusions}

We have used {\it Chandra} data for 15 early-type systems to constrain
the spectral and spatial properties of LMXBs in these galaxies. We have
found that once the most luminous ($> 10^{39}$ ergs s$^{-1}$) sources
are removed from the combined spectra of the sources, the spectra of the
sum of the sources are very similar values among the galaxies.
When all the galaxies are fit simultaneously with a power law spectral
model, the best-fit power law exponent is $\Gamma = 1.56 \pm 0.02$.
Even faint sources as dim as $10^{36}$ ergs s$^{-1}$ in the bulge of M31
have similar spectral properties as the more luminous sources. There was
no apparent difference in the spectral properties of LMXBs as a function
of galactic radius. Nor was there a significant difference in the spectral
properties of sources based on their presence within or outside a globular
cluster.

A significant number of sources with luminosities of $1-2 \times 10^{39}$
ergs s$^{-1}$ were found within the galaxies, and they exhibited
significantly softer spectral properties than the fainter sources. The
disk blackbody + power law model used to model their spectra is very
reminiscent of Galactic black hole X-ray binaries when they are in their
very high state. Their spectra were also quite different from ULXs
found within spiral galaxies. The simplest explanation of these sources
is that they are $\sim7-15$ M$_{\odot}$ accreting near their Eddington limit.
The spatial distribution of these sources is significantly more extended
than the optical light.

With rare exception, sources more luminous than $2 \times 10^{39}$
ergs s$^{-1}$ are absent from early-type galaxies. The number and
spatial distribution of the sources with fluxes corresponding to
$2 \times 10^{39}$ ergs s$^{-1}$ or greater if they are at the distance of the
galaxy is consistent with them being unrelated background/foreground
sources. The only exceptions to this seems to be two
$\sim 4\times 10^{39}$ ergs s$^{-1}$ sources found within globular clusters
of NGC~1399. Their spectra are also quite different than that of
a typical spiral galaxy ULX. Their presence within a globular cluster
suggests that globular clusters might harbor intermediate mass black holes
that are accreting at a few percent of their Eddington limit.

Finally, we have discussed how these constraints on the spectral properties
of LMXBs, especially the fainter ones, can lead to better constraints to
the luminosity, temperature, and metallicity of the hot gas within
early-type galaxies that contain little gas.
\acknowledgments

We thank the referee, Alexis Finoguenov, for many useful comments that
improved the quality of the manuscript.
J. A. I. was supported by {\it Chandra} Fellowship grant PF9-10009,
awarded through the {\it Chandra} Science Center. J. N. B.
acknowledges support from NASA grants GO0-1148 and GO1-2087.
The {\it Chandra} Science Center is operated by the Smithsonian Astrophysical
Observatory for NASA under contract NAS8-39073.


\begin{references}

\reference{}
Angelini, L., Loewenstein, M., \& Mushotzky, R. F. 2001, ApJ, 557, L35

\reference{}
Bauer, F. E., Brandt, W. N., Sambruna, R. M., Chartas, G., Garmire, G. P.,
Kaspi, S., \& Netzer, H. 2001, AJ, 122, 182

\reference{}
Blair, W. P., Fesen, R. A., \& Schlegel, E. M. 2001, AJ, 121, 1497

\reference{}
Blanton, E. L., Sarazin, C. L., \& Irwin, J. A. 2001, ApJ, 552, 106

\reference{}
Brown, B. A., \& Bregman J. N. 1998, ApJ, 495, L75

\reference{}
Canizares, C. R., Fabbiano, G., \& Trinchieri, G. 1987, ApJ, 312, 503

\reference{}
Caon, N., Capaccioli, M., \& D'Onofrio, M. 1994, A\&AS, 106, 109

\reference{}
Capaccioli, M., Held, E. V., \& Nieto, J.-L. 1987, AJ, 94, 1519

\reference{}
Chartas, G., \& Getman, K. 2002, ACISABS absorption profile

\reference{}
Colbert, E. J. M., \& Ptak, A. F. 2002, astro-ph/0204002

\reference{}
Davis, D. S., \& White, R. E. III 1996, ApJ, 470, L35

\reference{}
de Vaucouleurs, G. 1975, ApJS, 29, 193

\reference{}
Dickey, J. M., \& Lockman, F. J. 1990. ARA\&A, 28,215

\reference{}
Di Stefano, R., \& Kong, A. K. H. 2002, ApJ, submitted

\reference{}
Fabian A. C., \& Terlevich R. 1996, MNRAS, 280, L5

\reference{}
Finoguenov, A., \& Jones, C. 2002, ApJ, 574, 754

\reference{}
Gebhardt K., Pryor C., O'Connell R. D., Williams T. B., \& Hesser J. E. 
2000a, AJ, 119, 1268 

\reference{}
Gebhardt K., et al.\ 2000b, ApJ, 539, L13

\reference{}
Giacconi, R., Rosati, P., Tozzi, P., Nonino, M., Hasinger, G., Norman, C.,
Bergeron, J., Borgani, S., Gilli, R., Gilmozzi, R., \& Zheng, W.
2001, ApJ, 551, 624 (G01)

\reference{}
Goudfrooij, P., Hansen, L., Jorgensen, H. E., Norgaard-Nielsen, H. U.,
de Jong, T., \& van den Hoek, L. B. 1994, A\&AS, 104, 179
 van den Hoek, L. B.

\reference{}
Harris, W. E. 1991, ARA\&A, 29, 543

\reference{}
Immler S., Pietsch W., \& Aschenbach B. 1998, A\&A, 331, 601

\reference{}
Irwin, J. A., Sarazin, C. L., \& Bregman, J. N. 2002, ApJ, 570, 152

\reference{}
Jorgensen, I., Franx, M., \& Kjaergaard, P. 1995, MNRAS, 276, 1341

\reference{}
Kaastra J. S., \& Mewe R. 1993, A\&AS, 97, 443

\reference{}
Kim, D.-W., \& Fabbiano, G. 2002, astro-ph/0206369

\reference{}
King, A. R., Davies, M. B., Ward, M. J., Fabbiano, G., \& Elvis, M. 2001,
ApJ, 552, L109

\reference{}
Kormendy, J. 1984, ApJ, 286, 132

\reference{}
Kraft, R. P., Kregenow, J. M., Forman, W. R., Jones, C., \& Murray, S. S.
2001, ApJ, 560, 675

\reference{}
Kundu, A., Maccarone, T., \& Zepf, S. 2002, astro-ph/0206221

\reference{}
Liedahl D. A., Osterheld A. L., \& Goldstein W. H. 1995, ApJ, 438, L115

\reference{}
Maccarone, T., Kundu, A., \& Zepf, S. 2002, astro-ph/0210143

\reference{}
Makishima, K., et al.\ 2000, ApJ, 535, 632

\reference{}
Matsumoto, H., Koyama, K., Awaki, H., \& Tsuru, T., Loewenstein, M.,
\& Matsushita, K. 1997, ApJ, 482, 133

\reference{}
Mewe R., Gronenschild E. H. B. M., \& van den Oord G. H. J. 1985, A\&AS, 62, 197

\reference{}
Miller, M. C., \& Hamilton, D. P. 2002, MNRAS, 330, 232

\reference{}
Mizuno, T., Kubota, A., \& Makishima, K. 2001, ApJ, 554, 1282

\reference{}
Monet D., et al.,
1998, USNO-A V2.0, A Catalog of Astrometric Standards
(Flagstaff: U.S.\ Naval Observatory)

\reference{}
Mushotzky, R. F., Cowie, L. L., Barger, A. J., \& Arnaud, K. A. 2000,
Nature, 404, 459 (M00)

\reference{}
Nowak, M. 1995, PASP, 107, 1207

\reference{}
Peletier, R. F., Davies, R. L., Illingworth, G. D., Davis, L. E., \&
Cawson, M. 1990, AJ, 100, 1091

\reference{}
Perrett, K. M., Hanes, D. A., Butterworth, S. T., Kavelaars, J. J.,
Geisler, D., \& Harris, W. E. 1997, AJ, 113, 895

\reference{}
Randall, S. W., Sarazin, C. L., \& Irwin, J. A. 2002, ApJ, submitted

\reference{}
Roberts, T. P., \& Warwick, R. S. 2000, MNRAS, 315, 98

\reference{}
Ryden, B. S., Forbes, D. A., \& Terlevich, A. I. 2001, MNRAS, 326, 1141

\reference{}
Sarazin, C. L., Irwin, J. A., \& Bregman, J. N. 2000, ApJ, 544, L101

\reference{}
Sarazin, C. L., Irwin, J. A., \& Bregman, J. N. 2001, ApJ, 556, 533

\reference{}
Swartz, D. A., Ghosh, K. K., Suleimanov, V., Tennant, A. F., \& Wu, K.
2002, ApJ, submitted (astro-ph/0203487)

\reference{}
Soria, R., \& Wu, K. 2002, A\&A, 384, 99

\reference{}
Tanaka, Y., \& Lewin, W. H. G. 1995, in X-Ray Binaries, ed. W. H. G. Lewin,
J. van Paradijs, \& E. P. J. van den Heuvel (Cambridge: Cambridge Univ. Press),
126

\reference{}
Tonry, J. L., Dressler, A., Blakeslee, J. P., Ajhar, E. A., Fletcher, A., B.,
Luppino, G. A., Metzger, M. R., \& Moore, C. B. 2001, ApJ, 546, 681

\reference{}
van den Bergh, S. 2000, The Galaxies of the Local Group,
(Cambridge University Press, Cambridge)

\reference{}
van den Heuvel, E. P. J., Bhattacharya, D., Nomoto, K., \& Rappaport, S. A.
1992, A\&A, 262, 97

\reference{}
Vink J. S., de Koter A., \& Lamers H. J. G. L. M. 2001, A\&A, 369, 574

\reference{}
White, R. E. III 2001, astro-ph/0111293

\end{references}
\end{document}